\documentclass[nohyperref]{article}

\usepackage{microtype}
\usepackage{graphicx}
\usepackage{subfigure}
\usepackage{booktabs} 

\usepackage{hyperref}

\usepackage[accepted]{icml2022_TAGML}

\usepackage{amsmath}
\usepackage{amssymb}
\usepackage{mathtools}
\usepackage{amsthm}

\usepackage[capitalize,noabbrev]{cleveref}

\theoremstyle{plain}
\newtheorem{theorem}{Theorem}[section]

\theoremstyle{definition}

\theoremstyle{remark}

\usepackage[textsize=tiny]{todonotes}

\newcommand{\mb}[1]{\boldsymbol{#1}}
\newcommand{\bm}[1]{\mathbf{#1}}
\newcommand{\br}[1]{\left( #1 \right)}
\newcommand{\brs}[1]{\left[ #1 \right]}
\newcommand{\brangle}[1]{\left\langle #1 \right\rangle}
\newcommand{\brc}[1]{\left\{ #1 \right\}}

\newcommand{\R}{\mathbb{R}}

\newcommand{\norm}[1]{\left\lVert#1\right\rVert_2}
\newcommand{\T}{^\top}

\newcommand{\defeq}{\vcentcolon=}
\newcommand{\tr}{\mathrm{tr}}

\DeclareMathOperator*{\argmin}{arg\,min}

\graphicspath{{figures/}}

\renewcommand{\S}{\mathbb{S}}

\newcommand{\datadensitysphere}{q(\mb{z})} 
\newcommand{\modeldensitysphere}{p(\mb{z}; \mb\beta)}  

\newcommand{\modeldensityspheresample}{p(\mb{z}_i; \mb\beta)}  

\newcommand{\datascoresphere}{\score{q}(\mb{z})}
\newcommand{\modelscoresphere}{\score{\modeldensitysmall}(\mb{z})}
\newcommand{\modelscorespheresample}{\score{\modeldensitysmall}(\mb{z}_i)}

\newcommand{\manifold}{M}
\newcommand{\manifoldboundary}{\partial M}
\newcommand{\compactmanifold}{M_C}

\newcommand{\truncatedmanifold}{M_T}
\newcommand{\truncatedmanifoldboundary}{\partial M_T}

\newcommand{\manifoldd}{d_M}

\newcommand{\manifoldgrad}{\nabla_M}
\newcommand{\euclideangrad}{\nabla_E}

\newcommand{\dxj}{\partial_{x_j}}

\newcommand{\datadensity}{q(\mb{x})} 

\newcommand{\modeldensity}{p(\mb{x}; \mb\beta)}  
\newcommand{\modeldensitysmall}{p_{\mb\beta}}

\newcommand{\score}[1]{\mb{\psi}_{#1}}  

\newcommand{\datascore}{\score{q}(\mb{x})}
\newcommand{\modelscorex}{\score{\modeldensitysmall}(\mb{x})}

\newcommand{\obj}[1]{J_{\mathrm{\scriptscriptstyle #1}}(\mb\beta)}

\newcommand{\sampleobj}[1]{\hat{J}_{\mathrm{\scriptscriptstyle #1}}(\mb\beta)}

\icmltitlerunning{Score Matching for Truncated Density Estimation on a Manifold}

\begin{document}

\twocolumn[
\icmltitle{Score Matching for Truncated Density Estimation on a Manifold}



\icmlsetsymbol{equal}{*}

\begin{icmlauthorlist}
\icmlauthor{Daniel J. Williams}{yyy}
\icmlauthor{Song Liu}{yyy}
\end{icmlauthorlist}

\icmlaffiliation{yyy}{Department of Mathematics, University of Bristol, UK}

\icmlcorrespondingauthor{Daniel Williams}{daniel.williams@bristol.ac.uk}

\icmlkeywords{Machine Learning, ICML, Density Estimation, Parameter Estimation}

\vskip 0.3in
]



\printAffiliationsAndNotice{}  

\begin{abstract}
When observations are truncated, we are limited to an incomplete picture of our dataset. Recent methods propose to use score matching for truncated density estimation, where the access to the intractable normalising constant is not required. We present a novel extension of truncated score matching to a Riemannian manifold with boundary. Applications are presented for the von Mises-Fisher and Kent distributions on a two dimensional sphere in $\R^3$, as well as a real-world application of extreme storm observations in the USA. In simulated data experiments, our score matching estimator is able to approximate the true parameter values with a low estimation error and shows improvements over a naive maximum likelihood estimator.
\end{abstract}

\section{Introduction}
\label{intro}

Performing density estimation over the domain of a manifold such as the surface of the Earth, using methods that rely on a Euclidean data support, may yield inaccurate results due to the curvature of the planet's surface. This inaccuracy will be exacerbated when the region is significantly large, as the effect of the curvature becomes more pronounced. Earth is a manifold, and Euclidean-based methods do not account for this different shape.

Data collected in a geographic setting is likely to encounter a host of artificially occurring boundaries in the form of national or continental boundaries. For example, a country's borders may restrict data collection past an arbitrary stopping point, but the actual geographic structure of the Earth is unlikely to change beyond this point. Countries may only measure events that happen within their border, but not the phenomena which occur outside of their borders, for example in neighbouring countries or the ocean. Therefore it is likely that a dataset is truncated if the dataset originates from data collection confined within one country, making density estimation difficult. Regions defined by countries naturally span large areas of the Earth, making the curvature relevant to the task, exacerbating the difficulty for density estimation. 

The primary issue with estimating a truncated density is the evaluation of the normalising constant of a statistical model, which is often intractable in such a scenario. Score matching is a statistical estimation method which bypasses the evaluation of this normalising constant, making it suitable for estimating truncated densities. Score matching has seen many developments in recent years \citep{scorematching1, scorematching2}, and more broadly, unnormalised modelling methods such as score matching has seen many applications such as hypothesis testing \citep{ksd, kernelgoodness, Wu2022}, Markov random fields for image analysis \citep{li2009}, and independent component analysis \citep{teh2003}. Recent work has proposed a generic score matching objective to estimate a truncated density function \citet{Yu2021,song}. Methodology also exists for estimating a non-truncated statistical model on a manifold \citep{mardia2016}. However, no known score matching variant currently exists which estimates a truncated density on a manifold. In this work, we propose an efficient estimator that estimates such a truncated statistical model without requiring access to the normalising constant. 



\section{Background}

\subsection{Riemannian Manifold}

Firstly, let us define some notation to distinguish between manifold space and Euclidean space. Let $\manifold$ be any \textit{compact oriented Riemannian manifold}, and $\mb{z} \in \manifold$ is a $\manifoldd$-dimensional vector in \textit{local} coordinates. $\manifold$ is \textit{embedded} into the $d$-dimensional Euclidean space, $\R^d$, whose coordinates are denoted as $\mb{x} \in \R^d$.

Let $u(\mb{z})$ and $v(\mb{z})$ be two continuous twice differentiable real-valued functions on $\manifold$, with $\tilde{u}(\mb{x})$ and $\tilde{v}(\mb{x})$ being extensions of $u(\mb{z})$ and $v(\mb{z})$ to a neighbourhood $N \subset \R^m$ of $\manifold$. 

The gradient of $u(\mb{z})$ with respect to $\mb{z}$ on $\manifold$ is denoted by $\manifoldgrad u(\mb{z})$, and we denote its Euclidean counterpart as $\euclideangrad \tilde{u}(\mb{x}) = \nabla_{\mb{x}} \tilde{u}(\mb{x})$, for which
\begin{equation}
\manifoldgrad u(\mb{z}) = \bm{P} \euclideangrad \tilde{u}(\mb{x}),
\label{eq:manifoldgrad}
\end{equation}
where $\bm{P}$ is the $m \times m$ orthogonal projection matrix onto the $m$-dimensional tangent hyperplane to $M$. Additionally, we define the Laplace-Beltrami operator, which is given by
\begin{equation}
\Delta_{M} u=\tr\left\{\bm{P} \nabla_{\mb{x}} \cdot \left(\bm{P} \nabla_{\mb{x}} \tilde{u}\right)\right\},
\label{eq:lbo}
\end{equation}
and the manifold inner product, which is defined as
\begin{equation}
\langle u,v \rangle_M = \langle \bm{P}\tilde{u}, \bm{P}\tilde{v}\rangle.
\label{eq:innerprod}
\end{equation}

Finally, Stokes' theorem enables the following integration by parts \citep{stewart2020calculus}.

\begin{theorem} 
Let $f_1$ and $f_2$ be continuous twice differentiable functions on a compact Riemannian manifold $M$, then
\begin{equation}
\begin{split}
\oint_{\manifoldboundary} f_1(\mb{z}) &\frac{\partial \manifoldgrad f_2(\mb{z})}{ \partial \mathbf{n}} \manifoldboundary = \int_{\manifold} (\Delta_M f_2(\mb{z}))f_1(\mb{z})d\mb{z} \\
&+\int_{\manifold} \langle \manifoldgrad f_1(\mb{z}),   \manifoldgrad f_2(\mb{z})\rangle_M d\mb{z}.
\end{split}
\label{eq:stokes}
\end{equation}
where $\bm{n}$ is the unit normal vector.
\label{theorem:stokes}
\end{theorem}

See, e.g., \citet{manifolds} or \citet{jost2008} for a comprehensive overview of Riemannian manifolds.

\subsection{Score Matching}

Firstly, let us define a parametric statistical model as
\[
\modeldensitysmall = p(\mb{x};\mb\beta) = \frac{\bar{p}(\mb{x};\mb{\beta})}{Z(\mb\beta)}, \;\; Z(\mb{\beta}) = \int_{V} \bar{p}(\mb{x};\mb{\beta})d\mb{x},
\]
for some domain $V$ (e.g. $V = \R^d$), and where $p(\mb{x};\mb\beta)$ is the probability density function (pdf). We use this model to approximate an unknown data density function $q(\mb{x})$, where $\mb\beta$ is the parameter which would like to estimate. In many cases, the normalising constant, $Z(\mb\beta)$, is intractable, therefore classical methods such as maximum likelihood estimation (MLE) are unsuited. 

It is possible that probability density functions on $\manifold$ are defined in terms of either Euclidean coordinates ($\modeldensity$) or local coordinates ($\modeldensitysphere$).  We assume that the true data density is given by $\datadensitysphere$, defined in terms of local coordinates. 

\paragraph{Classical Score matching}

Score matching \citep{scorematching1} minimises
\begin{equation}
\obj{SM} = \int_V \datadensity\norm{\datascore - \modelscorex}^2d\mb{x},
\label{eq.0}
\end{equation}
since the \textit{score functions} $\datascore = \nabla_{\mb{x}} \log q(\mb{x})$ and $\modelscorex = \nabla_{\mb{x}} \log p(\mb{x};\mb{\beta})$ do not depend on the normalising constant. 

\paragraph{Generalised Score Matching}

Originally proposed to extend score matching to the non-negative domain, i.e. $\R^d_+$, generalised score matching \citep{scorematching2, yu2018sm, yu2019sm, yu2021sm}, weights the score matching objective with a weighting function, $g(\mb{x}) > 0$. This new objective is given by 
\begin{equation}
\obj{GSM} = \int_V q(\mb{x})g(\mb{x})\norm{\datascore - \modelscorex}^2d\mb{x}.
\label{eq:of1}
\end{equation}
Under some mild regularity conditions, the objective function in \eqref{eq:of1} can be written as
\begin{equation}
\begin{split}
\obj{GSM} = \int_{V}&q(\mb{x})\Big(g(\mb{x})\big[ \modelscorex^2 + 2\tr\{\nabla_{\mb{x}}\modelscorex\}\big]\\ 
&+ 2\brangle{\nabla_{\mb{x}} g(\mb{x}), \modelscorex}\Big)d\mb{x} + \text{const},
\end{split}
\label{eq:of2}
\end{equation}
where $\nabla_{\mb{x}}\modelscorex$ denotes the hessian of $ \log p(\mb{x}; \beta)$. 
This formulation is derived using integration by parts (see Theorem 3, \citet{positivegraphical}). Now the objective no longer depends on the density of the unknown data density $q(\mb{x})$, and thus the estimator of $\mb{\beta}$ is obtained by minimising a sample version of \cref{eq:of2}.

\paragraph{Truncated Score Matching}

In a \textit{truncated} setting, consider a bounded open subspace $V \subset \R^d$, where the boundary of the subspace is denoted by $\partial V$. We observe only $\mb{x}\in V$, and the integration required to calculate $Z(\mb\beta)$ is intractable. Therefore score matching, which does not require knowledge of $Z(\mb\beta)$, is a natural choice of method for estimation. However, the conditions which allowed the score matching objective \eqref{eq:of1} to be written tractably as \eqref{eq:of2} are no longer valid for the truncated domain $V$. Instead, \citet{song} introduced an extension which enforced a boundary constraint on the weighting function, given by $g(\mb{x}') = 0$ for all $\mb{x}' \in \partial V$. By design of $g$ for which this holds, the integration by parts required for a tractable score matching objective now holds for the truncated case.

\citet{song} proposed the use of a specific family of $g$ functions given by distance metrics. For example, the Euclidean distance given by $g(\mb{x}) = \min_{\mb{\tilde{x}} \in \partial V} \norm{\mb{x} - \mb{\tilde{x}}}$. Distance functions naturally satisfy the criteria for $g(\mb{x})$.

\paragraph{Manifold Score Matching}

Let us denote $\compactmanifold$ as a \textit{closed} oriented Riemannian manifold, such that $\compactmanifold$ is compact and oriented \textit{without} any edges or boundaries. Consider, for example, the sphere $\S^2$.
\citet{mardia2016} proposed a variant of score matching that is defined on $\compactmanifold$. Let $\mb{z} \in \compactmanifold$ and $\mb{z} \sim \datadensitysphere$. The \textit{manifold} score matching objective is given by
\begin{equation}
\obj{MSM} = \int_{\compactmanifold} \datadensitysphere \norm{\datascoresphere - \modelscoresphere}^2 d\mb{x},
\label{eq:ofM}
\end{equation}
where $\datascoresphere \defeq \manifoldgrad \log \datadensitysphere$ and $\modelscoresphere \defeq \manifoldgrad \log \modeldensitysphere$ are the score functions. Like preceding score matching methods, \cref{eq:ofM} is intractable in its current state due to the unknown quantity $\datascoresphere$. \citet{mardia2016} made use of \cref{theorem:stokes} to rewrite the objective, noting that since $\compactmanifold$ is closed, there is no need for boundary terms and the left hand side of \cref{eq:stokes} is zero. Hence, \cref{eq:ofM} can be written (up to a constant) as 
\begin{equation}
\begin{split}
\obj{MSM} = \int_{\compactmanifold} \datadensitysphere&\big[\langle \modelscoresphere, \modelscoresphere \rangle_M \\
&+2 \Delta_M \log \modeldensitysphere] d\mb{z}.
\end{split}
\label{eq:mardia}
\end{equation}

This can be written as an expectation, which does not involve evaluating the unknown data density $\datadensitysphere$. See Section 3, \citet{mardia2016} for details. This step is an analogue to using integration-by-parts in classical score matching to derive a tractable objective function.

Expectations in the objective function can be replaced with their Monte Carlo approximations to yield a tractable equation for use in density estimation.

\subsection{Spherical Domain} 

Motivated by geographical data on the surface of the Earth, we primarily consider the unit $(d-1)$-dimensional hypersphere embedded in $\R^d$, i.e. where $\manifold = \S^{d-1}$, given by
\[
\S^{d-1} \defeq \{\mb{x} \in \R^d \;|\; \|\mb{x}\|_2 = 1 \}.
\]
On the 2D sphere, Euclidean coordinates have components $\mb{x} = (x_1, x_2, x_3) \in \R^3$, whereas spherical coordinates have components $\mb{z} = (\theta, \phi) \in S^2$, where $\theta \in [0, \pi)$ is the colatitude and $\phi \in [0, 2\pi]$ is the longitude.  Conversion between spherical and Euclidean coordinates is given by
\[
\mb{x} = (r\cos\phi, r\sin\phi\cos\theta, r\sin\phi\sin\theta),
\]
and conversion from Euclidean coordinates to spherical coordinates is given by
\begin{equation}
\mb{z} = (\arccos\br{x_1/r}, \arctan (x_3/x_2)),
\label{eq:spherex}
\end{equation}
where $r = (x_1^2 + x_2^2 + x_3^2)^{1/2}$.
The projection matrix in this case is $\bm{P} = \bm{I}_3 - \mb{x}\mb{x}\T$ \citep{mardia2016}. Using this projection matrix, we can expand the Laplace-Beltrami operator (\cref{eq:lbo}) on $u(\mb{z})$ as follows
\begin{align}
&\mb\Delta_M  u(\mb{z}) = \tr\{\bm{P} \euclideangrad\T (\bm{P} \euclideangrad \tilde{u}(\mb{x})) \}\notag \\
	&\;= \tr\{\bm{P} \euclideangrad\T [(\bm{I}_3 - \mb{x}\mb{x}\T)  \tilde{u}(\mb{x})] \} \notag \\
    &\;= \tr\brc{\bm{P}\brs{\euclideangrad\T (\bm{I}_3 - \mb{x}\mb{x}\T)  \tilde{u}(\mb{x}) + (\bm{I}_3 - \mb{x}\mb{x}\T)\euclideangrad \tilde{u}(\mb{x})}} \notag \\
    &\;= \tr\brc{\bm{P} \brs{-2  \tilde{u}(\mb{x}) \mb{x}\T + \bm{P}\euclideangrad  \tilde{u}(\mb{x})}}.
\end{align}
In implementation, we assume an independent and identically distributed dataset given by either 
\[
\{\mb{z}_i\}^n_{i=1} = \{(\theta_i, \phi_i)\}^n_{i=1},
\] 
drawn from $\datadensitysphere$, which can be mapped to
\[
\{\mb{x}_i\}^n_{i=1} = \{(x_{i,1}, x_{i,2}, x_{i,3})\}^n_{i=1}
\] 
without loss of generality.

\subsubsection{Spherical Distributions} \label{sec:spheredists}

We consider two primary probability distributions on $\S^{d-1}$ in $\R^d$: the Kent distribution and the von-Mises Fisher distribution. See Chapter 9, \citet{directional} for an overview of spherical distributions. Note that these spherical distributions are defined in terms of Euclidean coordinates, $\mb{x} \in \R^d$.

\paragraph{Kent Distribution}

The PDF for the Kent distribution is
\[
\modeldensity = \frac{1}{Z(\kappa, \mb\alpha)} \exp \{ \kappa \mb\mu^\top \mb{x} + \sum_{j=1}^{d-1}\alpha_j (\mb\gamma_j^\top\mb{x})^2 \},
\]
for $\mb{x} \in \R^d$, $\mb\beta = (\mb\mu, \mb\gamma_1, \mb\gamma_2, \kappa, \mb\alpha)$, and where $Z(\kappa, \mb\alpha)$ is the normalising constant that ensures $\modeldensity$ integrates to one \citep{fisherbingham}. Here, $\mb\mu$ is the mean direction, whilst $\mb\gamma_j$ are the axes which determine the orientation of the probability contours. The concentration parameter, $\kappa$, and the ovalness parameters, $\alpha_j$, determine how elliptical the distribution is \citep{directional}. 

The Kent distribution is analogous to the Multivariate Normal (MVN) distribution on $\R^d$, and has similar properties, such as $\kappa$, $\alpha_j$, and $\mb\gamma_j$ controlling the shape of the data distribution, similar to the role of the covariance matrix of the MVN, and the mean direction $\mb\mu$ being similar to the mean of the MVN. The Kent distribution is sometimes referred to as the Fisher-Bingham distribution.

\paragraph{von Mises-Fisher Distribution}

The von Mises-Fisher distribution is a special case of the Kent distribution where $\alpha = 0$. It has PDF
\[
\modeldensity = \frac{1}{Z(\kappa)} \exp \{ \kappa \mb\mu^\top \mb{x}\},
\]
for $\mb{x} \in \R^d$ and $\mb\beta = (\mb\mu, \kappa)$. Here, $Z(\kappa)$ is the normalising constant, $\mb\mu$ is the mean direction and $\kappa$ is the concentration parameter \citep{directional}; all defined the same as in the Kent distribution. 

The von Mises-Fisher distribution can also be considered as analogous to the MVN with mean $\mb\mu$ and covariance matrix $\mb\Sigma = \kappa^{-1}\bm{I}_d$.

\section{Truncated Manifold Score Matching}

Let us now consider a compact oriented manifold \textit{with boundary} as $\truncatedmanifold$, where the boundary is denoted as $\truncatedmanifoldboundary$. For example, the upper hemisphere of a two-dimensional sphere, $\S^2$, is given by
\[
\mathbb{H}^+ = \{\mb{z} \in \S^2 \;|\; z_2 > 0\}
\]
and has a boundary, $\truncatedmanifoldboundary$, given by the equator of $\S^2$. Instead of observing data that lie on the full $\compactmanifold$, we instead observe a subset on $\truncatedmanifold$, i.e. the support of $\datadensitysphere$ is the \textit{truncated} domain $\truncatedmanifold$. In this case, the assumption by \citet{mardia2016} that the boundary term in \cref{eq:stokes} is zero no longer holds.

\subsection{Truncated Score Matching Objective on a Manifold}

The aim of this section is to extend the approach of \citet{mardia2016}, using methods of truncated score matching described in \citet{song}, to combine truncated and manifold score matching, which we entitle Truncated Manifold Score Matching (TMSM). Similar to previous score matching implementations, we propose to minimise the expected squared distance between the score functions of the statistical model and the data,
\begin{equation}
\obj{TMSM} \defeq \int_{\truncatedmanifold} \datadensitysphere g(\mb{z}) \norm{\datascoresphere - \modelscoresphere}^2d\mb{x},
\label{eq:ofmain}
\end{equation}
where the key difference from \citet{mardia2016} is the integration over $\truncatedmanifold$ instead of $\compactmanifold$. Similar to generalised score matching \citep{yu2018sm,yu2019sm,yu2021sm}, $g(\mb{z})$ is a weighting function, for which we define the following boundary condition:
\begin{equation}
\lim_{\mb{z} \to \mb{z}'} g(\mb{z}) \datadensitysphere = 0
\label{eq:boundarycon}
\end{equation} 
where $\mb{z} \in \truncatedmanifold$, $\mb{z}' \in \truncatedmanifoldboundary$ and $\mb{z} \to \mb{z}'$ takes any point sequence converging to $\mb{z}'$ in account. This condition is key, but is easily verified by choice of $g$, which will be a focus in the coming sections. Along with \cref{theorem:stokes}, this is used to derive the following theorem. 

\begin{theorem}[Truncated Manifold Score Matching Objective]
Given a function $g$ for which \cref{eq:boundarycon} holds, then $\obj{TMSM}$ can be expressed as
\begin{align}
\obj{TMSM} = \int_{\truncatedmanifold} & \datadensitysphere g(\mb{z}) \langle \modelscoresphere, \modelscoresphere\rangle_M d\mb{z} \notag \\
+ &2\int_{\truncatedmanifold} \datadensitysphere g(\mb{z}) \Delta_M \log \modeldensitysphere d\mb{z} \notag\\ 
+ &2\int_{\truncatedmanifold} \datadensitysphere \langle \manifoldgrad g(\mb{z}), \modelscoresphere\rangle_M d \mb{z},
\label{eq:ofmainint}
\end{align}
where the integration is over $\truncatedmanifold$, $\langle \cdot, \cdot \rangle_M$ is the manifold inner product and $\Delta_M$ is the Laplace-Beltrami operator.
\label{thm:main}
\end{theorem}
For proof, see \cref{app:proof1}.
Provided that we have a set of samples, $\{\mb{z}_i\}_{i=1}^n \sim \datadensitysphere$, then using the law of large numbers, a sample version of \cref{eq:ofmainint} can be written as 
\begin{equation}
\begin{split}
\sampleobj{TMSM} = \frac{1}{n}\sum_{i=1}^n & g(\mb{z}_i) \langle \modelscorespheresample, \modelscorespheresample\rangle_M \\
&+ \frac{2}{n}\sum_{i=1}^n g(\mb{z}_i) \Delta_M \log \modeldensityspheresample	  \\
&+ \frac{2}{n}\sum_{i=1}^n \langle  \manifoldgrad g(\mb{z}_i),  \modelscorespheresample\rangle_M ,
\end{split}
\label{eq:ofmainsample}
\end{equation}
provided that $n$ is sufficiently large, and where $\modelscorespheresample = \manifoldgrad \log \modeldensityspheresample$. The TMSM estimator is given by
\[
\mb{\hat{\beta}} \defeq \argmin_{\mb\beta} \sampleobj{TMSM}.
\]

Although the weighting function, $g(\mb{z})$, has not yet been specified, we only require a $g$ such that the condition for which $g(\mb{z}') = 0$ for all $\mb{z}' \in \manifoldboundary$ holds, provided it is at least once differentiable. In \cref{sec:fns}, we propose two distance functions that naturally fulfill this criteria, and are suited for the sphere.

\subsection{Score Matching on Spherical Distributions}

The score matching approach detailed above is applicable to any manifold on which Stokes' theorem (\cref{theorem:stokes}) holds. This includes the unit $(d-1)$-hypersphere in $\R^d$ from which the probability distributions detailed in \cref{sec:spheredists} are defined. These distributions are defined in terms of Euclidean coordinates, i.e. $\modeldensity$. The TMSM objective requires specification of the score function, $\manifoldgrad \modelscoresphere$, its inner product with itself, $\langle \modelscoresphere, \modelscoresphere\rangle_M$, and the Laplace-Beltrami operator, $\Delta_M \log \modeldensitysphere$, all in terms of local coordinates. Making use of the relationship between these operations in Euclidean space and manifold space, via \cref{eq:innerprod,eq:manifoldgrad,eq:lbo}, we detail the required operations on the von-Mises Fisher and Kent distribution in this section.

\paragraph{Kent Distribution}

The score function and its derivative in Euclidean space can be calculated as
\begin{align}
\modelscorex &= \kappa \mb\mu + 2\sum^{d-1}_{j=1} \alpha_j \mb\gamma_j \odot (\mb\gamma_j^\top\mb{x}),\\ 
\euclideangrad \modelscorex &=  2\sum^{d-1}_{j=1} \alpha_j \mb\gamma_j \mb\gamma_j\T, \notag 
\end{align}

where $\odot$ represents elementwise product. Using the formulae for the gradient (\cref{eq:manifoldgrad}), inner product (\cref{eq:innerprod}), and Laplace-Beltrami operator (\cref{eq:lbo}), for the Kent distribution we have
\begin{align*}
\modelscoresphere &= \bm{P}\Big(\kappa \mb\mu + 2\sum^{d-1}_{j=1} \alpha_j \mb\gamma_j \odot (\mb\gamma_j^\top\mb{x})\Big), \\
\langle{\modelscoresphere, \modelscoresphere}\rangle_M &= (\bm{P} \kappa \mb\mu)\T(\bm{P} \kappa \mb\mu) \\
&\hspace{-20mm}+ 4 \norm{\bm{P} \brs{\alpha_1\cdot \mb\gamma_1 (\mb\gamma_1^\top\mb{x}) + \alpha_2\cdot\mb\gamma_2 (\mb\gamma_2^\top\mb{x})}}^2 \vphantom{\sum}, \\
\Delta_M \log \modeldensitysphere &= 2\tr\{\bm{P}\big(\alpha_1\cdot\mb\gamma_1 \mb\gamma_1\T \\
&\;\;+ \alpha_2\cdot\mb\gamma_2\mb\gamma_2\T\big) - \bm{P}\modelscorex \mb{x}\T \}.
\end{align*}

\paragraph{von Mises-Fisher Distribution}

The score function and its derivative in Euclidean space are given by
\begin{align}
\modelscorex  &= \kappa\mb\mu,  \\ 
\euclideangrad \modelscorex &= \bm{0}_{d \times d}, 
\end{align}
where $\bm{0}_{d \times d}$ denotes the $d \times d$ matrix of zeros.  Similarly, we have
\begin{align*}
\modelscoresphere &= \bm{P}(\kappa\mb\mu), \\
\langle{\modelscorex, \modelscorex}\rangle_M &= (\bm{P} \kappa \mb\mu)\T(\bm{P} \kappa \mb\mu), \\
\Delta_M \log \modeldensity &= -2\tr\{\bm{P}\modelscorex \mb{x}\T \}.
\end{align*}

\subsection{Choice of Weighting Function} \label{sec:fns}


The weighting function, $g(\mb{z})$, needs to satisfy the criteria that $g(\mb{z}') = 0$ $\forall \mb{z}' \in \truncatedmanifoldboundary$, i.e. the function takes zero at the boundary of the manifold. Choice of $g(\mb{z})$ is an important topic of study and will play a key role in the quality of the TMSM estimator.



We take inspiration from the methods proposed in \citet{song}, and aim to define a distance function. However, defining a distance function on the manifold is not as straightforward as it is for the Euclidean domain. Two propositions of distance functions for $\S^2$ are discussed in this section, which are then incorporated into a general weighting function.


\paragraph{Haversine distance}

For two points on $\truncatedmanifold \subset \S^2$, defined by $\mb{z} = (\phi, \theta) \in \truncatedmanifold$ and $\mb{z}' = (\phi', \theta') \in \truncatedmanifoldboundary$, the \textit{Haversine}, or \textit{great circle} distance, measures the geodesic distance between these two points along the surface of a sphere, and is given by
\[
\mathrm{dist}(\mb{z}, \mb{z}') \defeq 2 r \arcsin \br{\sqrt{u}},
\]
where
\[
u=\sin^{2}\left(\frac{\phi'-\phi}{2}\right)+\cos \left(\phi\right) \cos \left(\phi'\right) \sin ^{2}\left(\frac{\theta'-\theta}{2}\right).
\]


\paragraph{Projected Euclidean distance}

\begin{figure}[!t]
\centering
\includegraphics[width=.46\linewidth]{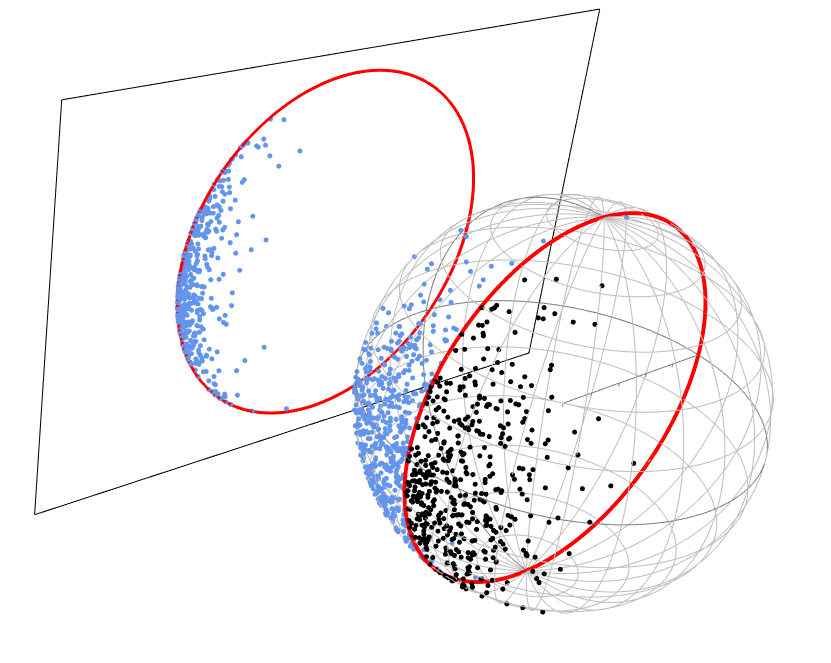}
\includegraphics[width=.46\linewidth]{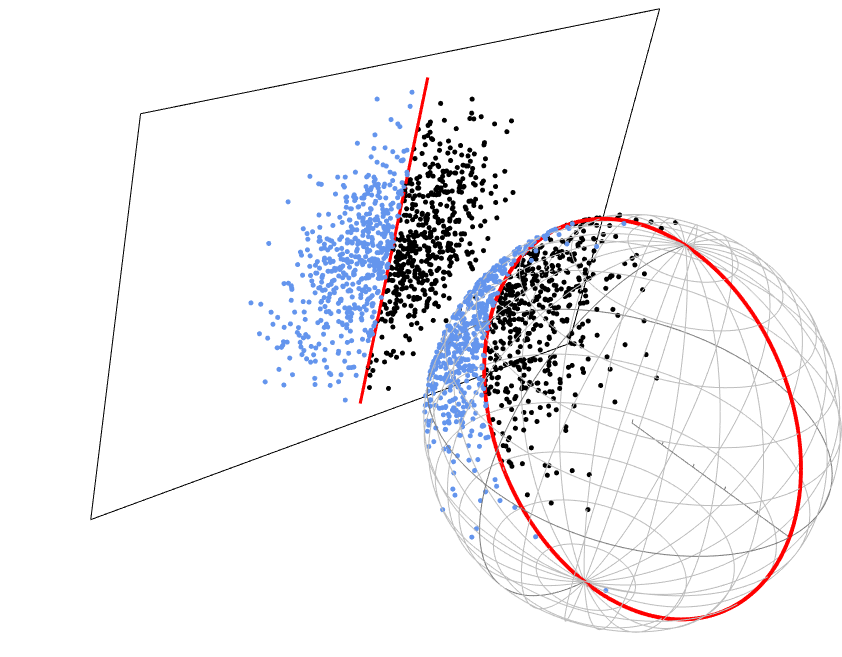}
\caption{Two visualisations of the Euclidean projection, projecting across a different axis (fixing a different coordinate to zero), with simulated points from a von Mises-Fisher distribution. A hemispherical boundary is visualised in red, with some unobserved points in black and observed points in blue. The projected Euclidean distance is calculated on the projected 2D plane.}\label{proj2:fig:projeuc}
\end{figure}

Points on the surface of the sphere, e.g. $\mb{x}, \mb{x}' \in \R^3$ can be projected on a two dimensional plane in $\R^2$ via setting the $j'$-th coordinate to a fixed constant (in practice, we can set it to zero). The corresponding two-dimensional Euclidean distance will be measured on this plane. See Figure \ref{proj2:fig:projeuc} for a visualisation of this method. 


Let $\mb{x}_e = \mathrm{Proj}(\mb{z})$, where $\mathrm{Proj}$ is a function that defines the projection from the sphere to the 2D Euclidean plane, then for two projected points $\mb{x}_e, \mb{x}'_e \in \R^2$, the distance function is given by
\[
\mathrm{dist}(\mb{z}, \mb{z}') \defeq \|\mathrm{Proj}(\mb{z}) - \mathrm{Proj}(\mb{z}')\|_2 = \|\mb{x}_e - \mb{x}'_e\|_2.
\]
The gradient is easily defined as
\[
\dxj \mathrm{dist}(\mb{z}, \mb{z}') = \frac{x_{e, j} - \tilde{x}_{e,j}}{\norm{\mb{x}_e - \mb{\tilde{x}}_e}}, \;\; \forall j \in \{1,\dots, d\} \backslash \{j'\}
\]
and $\partial_{x_{j'}}  \mathrm{dist}(\mb{z}, \mb{z}') = 0$, because the $j'$-th coordinate was fixed to a constant. Here, $x_{e, j}$ denotes the $j$-th element of $\mb{x}_e$, and $\tilde{x}_{e,j}$ denotes the $j$-th element of $\mb{\tilde{x}}_e$, which is the corresponding \textit{projection point} of $\mb{x}_e$; the closest point on the boundary to $\mb{x}_e$.

\begin{figure*}
\centering
\includegraphics[width=\linewidth]{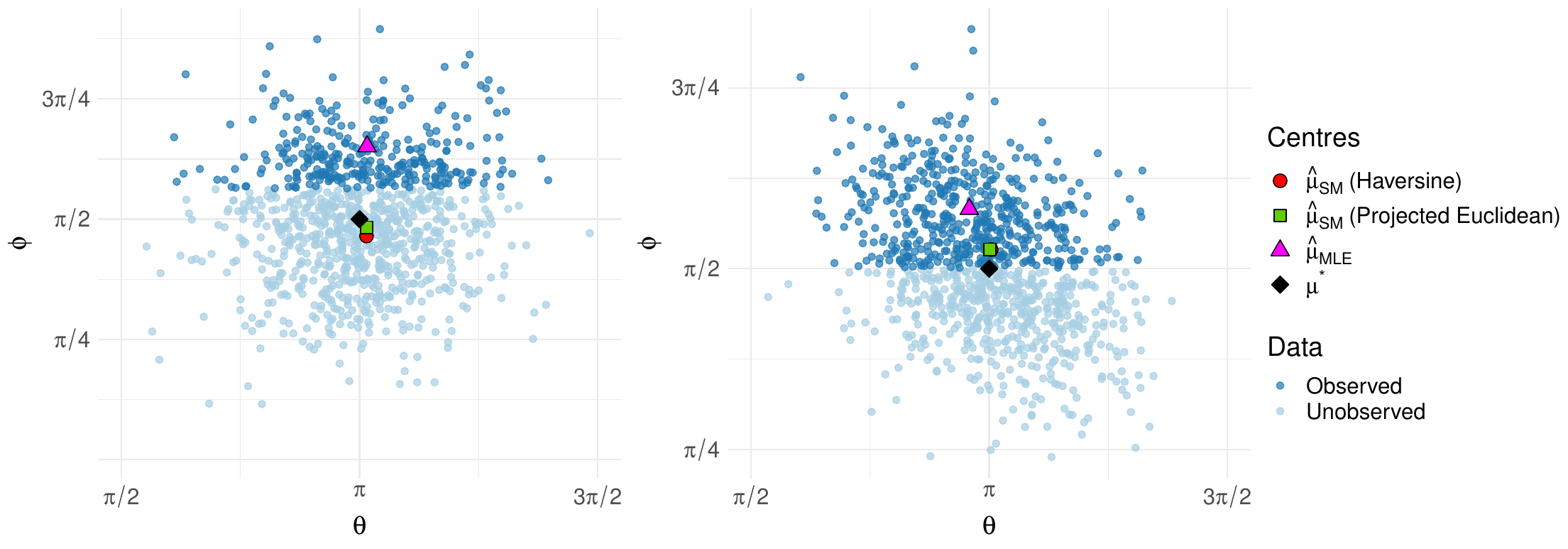}
\caption{Synthetic data experiments for TMSM. Von-Mises Fisher distribution (left) and Kent distribution (right).}
\label{fig:ex}
\end{figure*}

\begin{figure*}[!t]
\centering
\includegraphics[width=\linewidth]{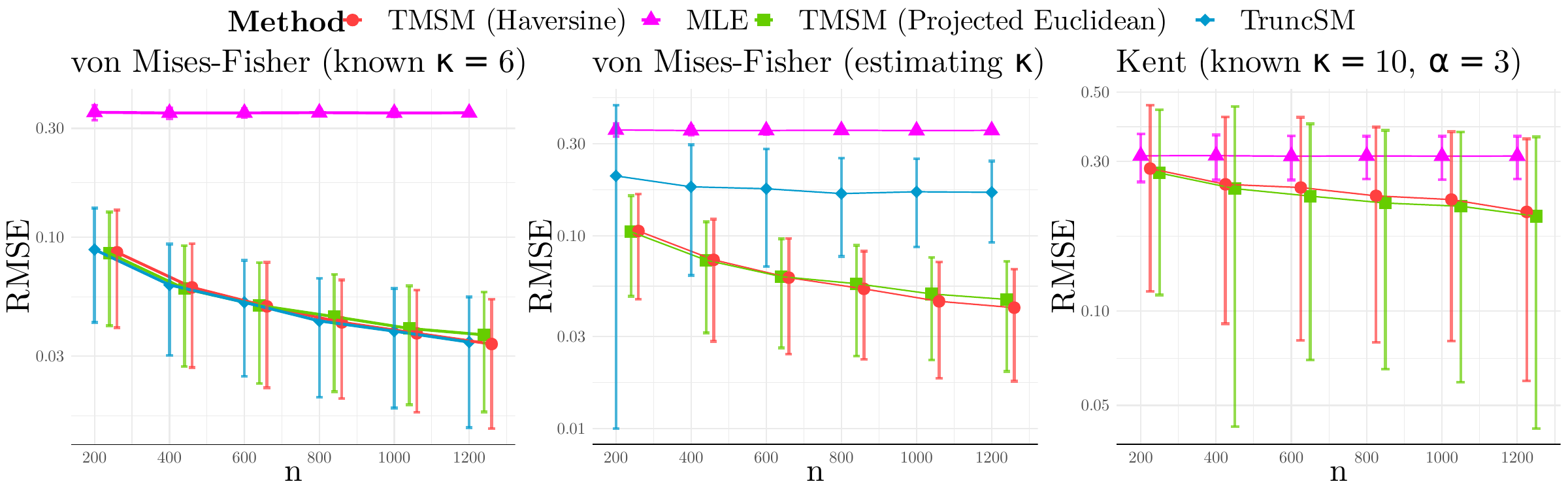}
\caption{Mean estimation error and standard deviation against sample size $n$, on a log scale, from 512 trials for each method.}
\label{fig:benchmarks}
\end{figure*}

\paragraph{Weighting function}
Provided that a distance function is defined, we define the weighting function by the shortest distance from a point $\mb{z}$ to the boundary $\truncatedmanifoldboundary$, given by 
\[
g(\mb{z}) = \min_{\mb{z}' \in \truncatedmanifoldboundary} \mathrm{dist}(\mb{z}, \mb{z}').
\]
In the case of the projected Euclidean distance, the boundary itself would also need to be projected into the Euclidean space. For any distance function, this weighting function naturally satisifes the boundary condition required by TMSM detailed in \cref{eq:boundarycon}.

\section{Experiments}

We focus on truncated domains on the sphere $\S^2$ in $\R^3$, as this manifold covers many important applications in geological or earth science applications where the datasets are collected from a truncated region on a the Earth. 

\subsection{Illustrative Example} \label{sec:sim}

We start with a synthetic data experiment, where data are simulated from a von Mises-Fisher distribution and Kent distribution using a rejection sampling algorithm, as recommended by \citet{simvmf}. For the first example, we simulate $n$ truncated samples from a von Mises-Fisher distribution with an unknown true mean direction $\mb\mu^\star = (\pi/2, \pi)$ and an unknown concentration parameter $\kappa^\star = 6$, where the truncation boundary is at a constant value of colatitude $\phi$. The aim of the experiment is to produce an estimate, $\mb{\hat\beta}^\star \defeq (\mb{\hat{\mu}}, \hat\kappa)$ of $\mb\beta^\star \defeq (\mb\mu^\star, \kappa^\star)$, the mean direction and the concentration parameter. \Cref{fig:ex} shows the results of this example using the two different weighting functions proposed in \cref{sec:fns}, compared to a naive MLE implementation which does not account for truncation. Both estimates of the mean direction lie roughly in the correct location.

The second example is a similar experiment, where samples are instead simulated from the Kent distribution with an unknown true mean direction $\mb\mu^\star = (\pi/2, \pi)$ and a known concentration parameter, $\kappa^\star = 10$, and ovalness parameter, $\alpha^\star = 3$.  The aim of this experiment is to produce an estimate $\mb{\hat{\mu}}$ of $\mb{\hat\mu}$, the mean direction only. \Cref{fig:ex} shows the results of this experiment. Similarly, the mean directions estimated appear close to the true value. 
 
These plots serve as an example of the full capability of the method; even with half of the observations missing, the estimates of the proposed methods are closer to the true mean direction than MLE.

\subsection{Estimation Accuracy} \label{sec:bench}

\begin{figure*}[!t]
\centering
\includegraphics[width=0.825\linewidth]{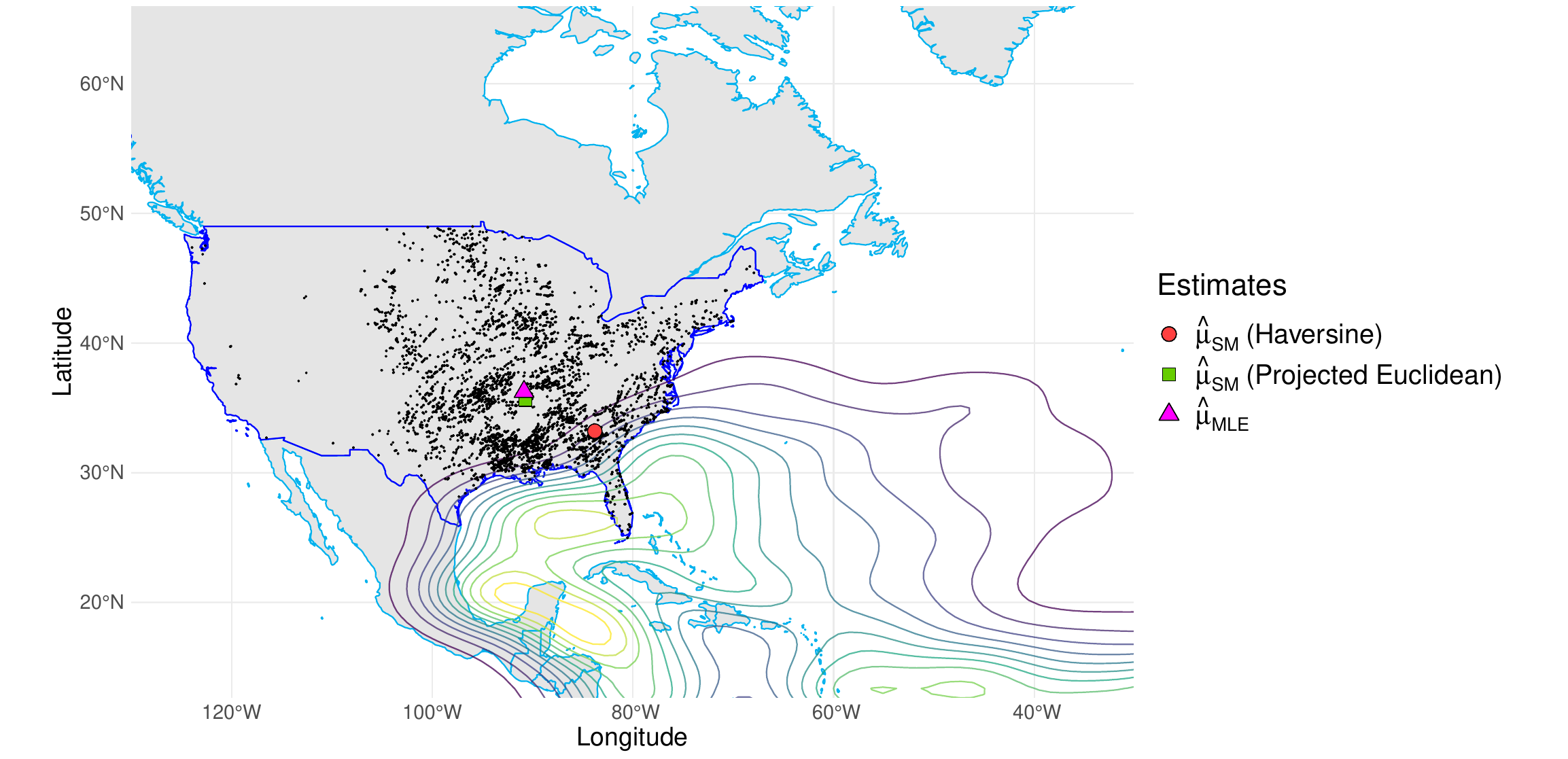}
\caption{Observed storms as recorded by the U.S. from 2017 to 2020, with a kernel density estimate contour of storm origin locations over the Atlantic ocean.} \label{proj2:fig:storms}
\end{figure*}

We repeat experiments, similar to those presented in \cref{sec:sim}, across different setups. For each trial, we simulate and truncate samples until we have $n$ truncated samples, simulated from either the von Mises-Fisher or Kent distribution. The truncation boundary in this case is the upper hemisphere, i.e. $\truncatedmanifold = \mathbb{H}^+$, where only datapoints for which $\phi > \pi / 2$ were observed. Across different values of sample size $n$, \cref{fig:benchmarks} shows the root mean squared error (RMSE), given by
\[
\mathrm{RMSE}(\mb{\hat\beta}, \mb\beta^\star) = \frac{1}{d}\sqrt{\sum^d_{j=1} (\hat{\beta}_j - \beta_j^\star)^2}.
\]
The three experiments are summarised as follows. First, we calculate the RMSE between the true mean direction, $\mb\mu^\star$, of the von Mises-Fisher distribution with a known concentration parameter $\kappa$, and the corresponding estimates $\mb{\hat\mu}$. The second experiment is the estimation of $\mb\beta^\star \defeq [\mb\mu^\star, \kappa^\star]\T$, and the RMSE is reported between $\kappa^\star$ and $\hat{\kappa}$. Finally, given a known concentration parameter $\kappa$ and ovalness parameter $\alpha$, the RMSE is calculated between the true mean direction, $\mb\mu^\star$ of the Kent distribution and the estimates $\mb{\hat\mu}$.

These estimates are compared across three methods: 
\textit{TruncSM} \citep{song}, with a multivariate Normal assumption\footnote{
Whilst the TMSM methods can estimate $\kappa$ directly, for \textit{TruncSM} we let $\mb\mu_{\text{MVN}} = \mb\mu_{\mb{z}}$, i.e. the mean of the MVN distribution is the mean direction of the von Mises-Fisher distribution in polar coordinates, and $\mb\Sigma_{\text{MVN}} = \kappa^{-1}\bm{I}_d$, i.e. the covariance matrix of the MVN is isotropic. In this case, we treat $\kappa^{-1}$ as the precision parameter and estimate it directly with \textit{TruncSM}.
}, 
our method Truncated Manifold Score Matching (TMSM), with either the Haversine or projected Euclidean distance, and MLE for a baseline. 

In all cases the estimation error decreases as sample size increases, indicating empirical consistency. Comparing the two distance functions for $g(\mb{z})$, both give near identical results across all benchmarks. Our method achieves comparable performance to \textit{TruncSM} in the most cases, and outperforms it when $\kappa$ is also estimated. This is also expected as \textit{TruncSM} incorrectly assumes a Euclidean domain.

\subsection{Storm Events in the U.S.}

Consider an application where the task is to estimate the probable locations of extreme storm events in the U.S., with dataset given by \citet{storm}. Whilst storms are not restricted to land, observations of these events only happen on land and within the country's borders. Furthermore, the large area spanned by the U.S. is significant enough that the curvature of its surface will affect any estimation, highlighting the need for considering density estimation on a truncated manifold.


We hypothesise that more storms are likely to occur on the East Coast of the U.S., and spread across the land from there. We assume there is a true storm `centre' where most storms will be spread out from beyond the country's eastern border \citep{earthclim}. 

Assuming the data can be modelled accurately by a von Mises-Fisher distribution, we produce an estimate of the mean direction, $\mb{\hat\mu}$. The estimates for both MLE and TMSM are given in Figure \ref{proj2:fig:storms}. Included in the figure is a kernel density estimate of storm points of origin, which are calculated via a post-storm analysis, after a storm is observed \citet{hurdat}.
Unsurprisingly, we see that $\mb{\hat\mu}$ given by MLE is at the center of the dataset, whereas the truncated manifold score matching estimate is further South East, closer to the storm origins. 
\section{Discussion}

The aim of this work was to derive an estimator for truncated densities on a generic manifold. We applied Stokes' theorem and a scaling function to obtain a tractable score matching objective, of which the estimation was possible via numerical minimisation. The method was tested on the two dimensional sphere in $\R^3$, where we showed a superior performance comparing to a Euclidean approximation using TruncSM. A further application was presented on severe storms in the USA, in which the score matching estimator gave a mean direction estimate lying on the East Coast, where the bulk of the storms originated. 

Our estimator should also be applicable to any manifold in which Stokes' Theorem holds, but applications have only been studied on the two dimensional sphere. Other complicated manifolds would require different distribution specification and different scaling functions. It is likely there exists a general purpose scaling function on a manifold that involves projection into a Euclidean domain, which could generalise this work further, but it is beyond the scope of this work.

\section*{Software and Data}

A fully packaged implementation in R, with examples and a tutorial, can be found at the GitHub repository here:
\url{https://github.com/dannyjameswilliams/truncsm}.

\section*{Acknowledgements}

We would like to thank Dr. Farhad Babaee for the generous help with the geometric algebra portion of this work. Additionally, we would like to thank the two anonymous reviews for their feedback. Daniel J. Williams was supported by a PhD studentship from the EPSRC Centre for Doctoral Training in Computational Statistics and Data Science (COMPASS).

\nocite{directional}
\nocite{manifolds}

\nocite{r}
\nocite{rdirectional}
\nocite{rrfast}
\nocite{rggplot2}

\bibliography{main}
\bibliographystyle{icml2022}

\newpage
\appendix
\onecolumn
\section{Proof of Theorem \ref{thm:main}} \label{app:proof1}
\begin{figure}[!t]
\centering
\includegraphics[width=0.7\linewidth]{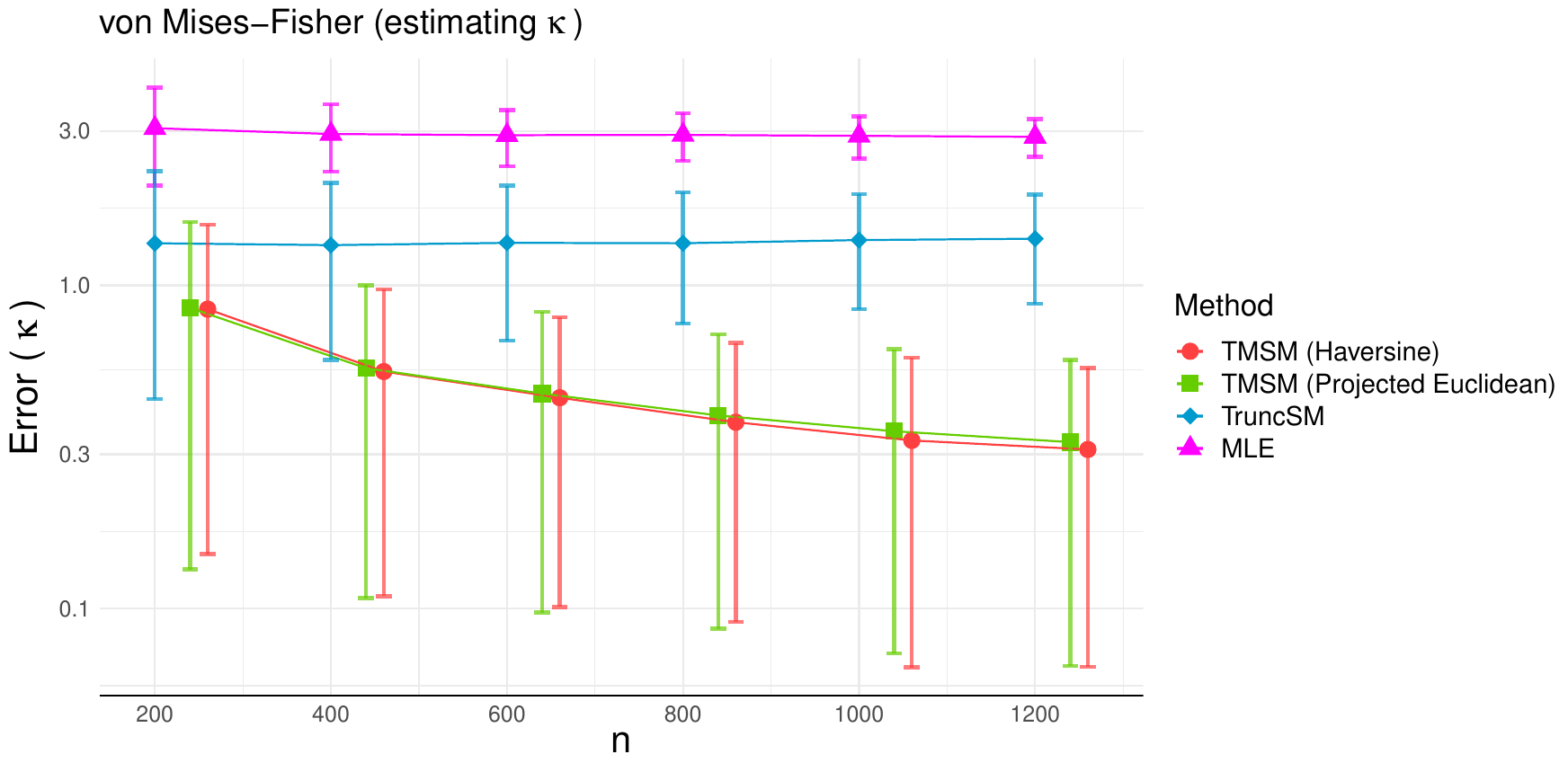}
\caption{Root mean squared error (with standard deviation) from 512 replicates between $\kappa^*=6$ and the estimated $\hat{\kappa}$ given by the different methods.} \label{fig:benchkappa}
\end{figure}

\begin{proof}
We can expand out \eqref{eq:ofmain} to give
\begin{align*}
\obj{TMSM} = \int_{\truncatedmanifold} \datadensitysphere g(\mb{z}) \Big[ \langle \modelscoresphere, \modelscoresphere \rangle_M - 2\langle \modelscoresphere, \datascoresphere\rangle_M \Big] d\mb{z} + C_q, 
\end{align*}
where $C_q = \int_{\truncatedmanifold} \datadensitysphere g(\mb{z}) \langle \manifoldgrad\log \datadensitysphere, \manifoldgrad\log \datadensitysphere \rangle_M d\mb{x}$. Note that we can write
\[
\int_{\truncatedmanifold} \datadensitysphere g(\mb{z}) \langle \modelscoresphere, \datascoresphere\rangle_M  d\mb{z} = \int_{\truncatedmanifold}  g(\mb{z}) \langle  \modelscoresphere, \manifoldgrad \datadensitysphere\rangle_M d\mb{z},
\]
due to $\manifoldgrad \log \datadensitysphere = \datadensitysphere^{-1} \manifoldgrad \datadensitysphere$, and therefore
\begin{equation*}
\obj{TMSM} = \int_{\truncatedmanifold} \datadensitysphere g(\mb{z}) \Big[ \langle \modelscoresphere, \modelscoresphere \rangle_M - 2\langle  \modelscoresphere, \manifoldgrad \datadensitysphere\rangle_M \Big] d\mb{z} + C_q.
\end{equation*} 
Now, noting that \cref{theorem:stokes} can be applied to $\truncatedmanifold$, and rearranged as 
\[
\int_{\truncatedmanifold} \langle \manifoldgrad f_1(\mb{z}),   \manifoldgrad f_2(\mb{z})\rangle_M d\mb{z} = \oint_{\truncatedmanifoldboundary} f_1(\mb{z}) \frac{\partial \manifoldgrad f_2(\mb{z})}{ \partial \mathbf{n}} \truncatedmanifoldboundary -  \int_{\truncatedmanifold} (\Delta_M f_2(\mb{z}))f_1(\mb{z})d\mb{z},
\]
and then letting $f_1 = g(\mb{z})\datadensitysphere$ and $f_2=\log \modeldensity$, on our manifold $\truncatedmanifold$ with boundary $\truncatedmanifoldboundary$, we have
\begin{align*}
\int_ {\truncatedmanifold}\langle \manifoldgrad [g(\mb{z})\datadensitysphere],   \modelscoresphere\rangle_M d\mb{z} &= \oint_{\truncatedmanifoldboundary} g(\mb{z})\datadensitysphere \frac{\partial \modelscoresphere}{ \partial \mathbf{n}} \truncatedmanifoldboundary \\
&\qquad -  \int_{\truncatedmanifold} (\Delta_M \log \modeldensitysphere)g(\mb{z})\datadensitysphere d\mb{z},
\end{align*}
where $\bm{n}$ is the unit normal vector. By \cref{eq:boundarycon}, the boundary term is equal to zero, leaving
\begin{align*}
\int_{\truncatedmanifold} \langle \manifoldgrad [g(\mb{z})\datadensitysphere],  \modelscoresphere\rangle_M d\mb{z} = -  \int_{\truncatedmanifold} (\Delta_M \log \modeldensitysphere)g(\mb{z})\datadensitysphere d\mb{z}.
\end{align*}
Expanding the left hand side via the chain rule and rearranging, gives
\begin{align*}
\int_{\truncatedmanifold} g(\mb{z}) \langle \manifoldgrad \datadensitysphere,  \modelscoresphere\rangle_M d\mb{z} 
&= -  \int_{\truncatedmanifold} (\Delta_M \log \modeldensitysphere)g(\mb{z})\datadensitysphere d\mb{z} \\
&\qquad - \int_{\truncatedmanifold} \datadensitysphere \langle \manifoldgrad g(\mb{z}),  \modelscoresphere\rangle_M d\mb{z}
\end{align*}
This can be substituted back into $\obj{TMSM}$ to obtain
\begin{align*}
\obj{TMSM} &= \int_{\truncatedmanifold} \datadensitysphere g(\mb{z}) \langle \modelscoresphere, \modelscoresphere\rangle_M d\mb{x} 
+ 2\int_{\truncatedmanifold} \datadensitysphere g(\mb{z})\Delta_M \log \modeldensitysphere d\mb{x} \\
& \qquad \qquad \qquad \qquad \qquad \qquad+ 2\int_{\truncatedmanifold} \datadensitysphere\langle \manifoldgrad g(\mb{z}), \modelscoresphere\rangle_M d\mb{x},
\end{align*}
which concludes the proof.
\end{proof}

\section{Notes on Benchmarks from Section \ref{sec:bench}} \label{app:truncsmexp}

In the first two experiments from Section \ref{sec:bench}, and the left two plots from Figure \ref{fig:benchmarks}, the task was estimating the mean direction $\mb\mu$ in the von Mises-Fisher distribution when the concentration parameter, $\kappa$, was known and unknown. Whilst the truncated manifold score matching methods can estimate $\kappa$ directly, for TruncSM we let $\mb\mu_{\text{MVN}} = \mb\mu_{\mb{z}}$, i.e. the mean of the MVN distribution is the mean direction of the von Mises-Fisher distribution in polar coordinates, and $\mb\Sigma_{\text{MVN}} = \kappa^{-1}\bm{I}_d$, i.e. the covariance matrix of the MVN is isotropic. In this case, we treat $\kappa^{-1}$ as the precision parameter and estimate it directly with the TruncSM formulation. Figure \ref{fig:benchkappa} shows a similar benchmark plot, giving the error for estimating $\kappa$, this plot is parallel to the middle-most plot in Figure \ref{fig:benchmarks}.

Whilst there is a clear difference between the errors for estimation from TruncSM and truncated manifold score matching in both Figures \ref{fig:benchmarks} and \ref{fig:benchkappa}, it is clear from the first experiment, where $\kappa$ is known, that the equality $\mb\Sigma_{\text{MVN}} = \kappa^{-1}\bm{I}_d$ holds, since the methods were very close in performance when $\kappa^{-1} = 1/6$ was fixed.. 


\end{document}